# Method for Assessing the Fidelity of Optical Diffraction Tomography Reconstruction Methods

**AHMED B. AYOUB,[1,*] THANH-AN PHAM,[2] JOOWON LIM,[1] MICHAEL UNSER,[2] AND DEMETRI PSALTIS[1]**

[1]Optics Laboratory, Ecole polytechnique fédérale de Lausanne (EPFL), 1015 Lausanne, Switzerland
[2]Biomedical Imaging Group, Ecole polytechnique fédérale de Lausanne (EPFL), CH-1015 Lausanne, Switzerland
* ahmed.ayoub@epfl.ch

**Abstract:** We use a spatial light modulator in a diffraction tomographic system to assess the accuracy of different refractive index reconstruction algorithms. Optical phase conjugation principles through complex media, allows us to quantify the error for different refractive index reconstruction algorithms without access to the ground truth. To our knowledge, this is the first assessment technique that uses structured illumination experimentally to test the accuracy of different reconstruction schemes.

1. Introduction

Label-free detection and characterization of microstructures and biological cells/specimens is the goal of many current detection techniques. Optical Diffraction Tomography (ODT) is an example of such quantitative characterization of biological specimens by reconstructing the 3D refractive index. Digital holography is used to record the complex field of projections taken at different illumination angles, from which the 3D refractive index reconstruction of structures can be evaluated. Several reconstruction methods have been proposed and demonstrated [1-20] but we cannot quantitatively assess their relative performance because the ground truth for the 3D distribution of biological 3D samples is not available. As a result phantom objects are used such as beads to evaluate the performance. In this paper we describe and experimentally demonstrate a method that provides a comparative metric for assessing the relative performance of reconstruction algorithms for arbitrary 3D objects without having access to their ground truth. The method we present is applicable to any 3D reconstruction technique that provides an estimate of the 3D index distribution of the sample. We will first describe the method in the following section and then present results comparing three commonly used ODT algorithms: Radon [10], Born [1] and Rytov [4-6].

2. Description of the Method

The distortion imposed on an optical field propagating through an inhomogeneous medium with negligible absorption can be undone if the transmitted field is holographically recorded and the phase conjugate reconstruction of the hologram is made to propagate backwards through the sample [add phase conjugation reference]. This is conveniently done in the optical domain by illuminating the recorded hologram with a plane wave counter-propagating to the plane wave used to record the hologram. When the incident beam illuminating the object is spatially modulated by a 2D pattern (an image), the field arriving at the hologram plane is a distorted version of the 2D illumination pattern. Through phase conjugation this distortion is removed and the field arriving back at the input plane is ideally an exact replica of the original image. Deviations from this ideal condition can occur due to limited spatial bandwidth, absorption or other losses in the optical path. Any imperfection in the holographic recording and play-back of the hologram (including speckle) can also contribute to deviations of the phase conjugate reconstruction from the original image projected through the sample. In a carefully designed optical system, we can generally obtain excellent phase conjugate reconstructions. The phase conjugate image is also strongly affected by any changes in the 3D object in the time between the recording of the hologram and the play-back. Therefore, if all other effects are negligible, then any distortions in the phase conjugate image can be attributed to changes in the object itself. This effect has been used for many applications including imaging through diffusing media, turbidity suppression in biological samples and imaging through turbid media [21 -24].

In our system we record the hologram of the 3D object on an sCMOS camera and the phase conjugation is performed digitally by computationally propagating the conjugate of the experimentally measured field though the 3D object whose index distribution has been measured through ODT. The system is shown schematically in Figure 1. If the ODT reconstruction and the digital wave propagation method are both accurate then we expect a faithful digital reconstruction of the image that was presented on the SLM. Any distortions in the digital reconstruction are attributed to inaccuracies of the ODT reconstruction algorithm. Measurement of the degree of distortion in the digital reconstruction provides a quantitative metric which we can use to compare ODT reconstruction algorithms.

The numerical method we use to digitally simulate light propagation through the sample is based on the Lippmann-Schwinger Equation (LSE) [25, 26]:

$$E(\boldsymbol{r}) = E_{inc}(\boldsymbol{r}) + \int G(\boldsymbol{r} - \boldsymbol{r}')E(\boldsymbol{r}')\eta(\boldsymbol{r}')\mathrm{d}\boldsymbol{r}',$$

where $E_{inc}$ and $E$ are the incident and total field respectively. G($\boldsymbol{r}$) denotes the Green function, $k = 2\pi n_m/\lambda$ is the optical wavenumber in the medium of refractive index $n_m$, and $\eta(\boldsymbol{r}) = k^2 \left(\frac{n(r)^2}{n_m^2} - 1\right)$ is the scattering cross-section of the sample of refractive index $n(\boldsymbol{r})$. Our numerical propagation is divided in two main steps methods. The discrete total field **E** in the region of interest (*i.e.,* which includes the sample) is computed as

$$\mathbf{E} = (\mathbf{I} - \mathbf{G}\,\boldsymbol{\eta})^{-1}\mathbf{E}_{\mathbf{inc}},$$

where $\boldsymbol{\eta}, \mathbf{E_{inc}}$ denote the discretized version of their continuous counterparts, and $\mathbf{G}$ denotes the discrete convolution with the Green function. In this work, we use the BiConjugate Gradients Stabilized Method to iteratively compute the matrix inverse.

The second step provides the total field at the sensors position

$$\mathbf{E_{scat}^{meas}} = \mathbf{G^{meas}} (\mathbf{E}\,\boldsymbol{\eta}) + \mathbf{E_{inc}^{meas}},$$

where $\mathbf{G^{meas}}$ and $\mathbf{E_{inc}^{meas}}$ denote the linear operation that yields the scattered field at the sensors position and the incident field at the sensors position respectively. The LSE method is expected to be superior to other linear and nonlinear methods such as the beam propagation method. Beyond the scalar assumption, there is no further approximation. The multiple scattering events (including the reflections) are fully accounted for as opposed to the beam propagation method.

The optical system shown in Fig. 1 used a diode pumped solid state (DPSS) 532 nm laser. The laser beam was first spatially filtered using a pinhole. A beam-splitter separated the input beam into a signal and a reference beam. The signal beam was directed to the sample at different angles of incidence using a reflective liquid crystal on silicon (LCOS) spatial light modulator (SLM) (Holoeye PLUTO VIS, pixel size: 8 μm, resolution: 1080x1920 pixels). Different illumination angles were obtained by recording blazed gratings on the SLM. The SLM was slightly rotated and calibrated first with an "alignment" grating (3 pixels) to eliminate the DC term from the back-focal plane of the illuminating objective and only the 1st order is allowed to pass through the objective so that we make sure that DC term is completely out of the focal-plane of the illuminating objective lens. After that, another "rotation" grating is superimposed on it to give the desired illumination angle in the object plane. In the experiments presented here, a blazed grating with a period of 25 pixels (200 μm) was rotated a full 360° with a resolution of 1 degree for a total of 361 projections (including normal incidence to be able to measure the shift in the k vectors with respect to it). Two 4f systems between the SLM and the sample permitted filtering of higher orders reflected from the SLM (due to the pixilation of the device) as well as magnification of the SLM projections onto the sample. Using a 100X oil immersion objective lens (OBJ1) with NA 1.4 (Olympus), the incident angle on the sample corresponding to the 200 μm grating was about 37°. The magnification of the illumination side was around 240 defined by the 4f systems we used before the sample. A third 4f system after the sample includes a 100X oil immersion objective lens (OBJ2) with NA 1.45 (Olympus). The sample and reference beams were collected on a second beam-splitter and projected onto a scientific complementary metal-oxide-semiconductor (sCMOS) camera (Andor Neo 5.5 sCMOS, pixel size: 6.5 μm, resolution: 2150 x 2650 pixels). The samples used were HCT-116 human colon cancer cells and Panc-1 human pancreas cancer cells which were cultured in McCoy 5A growth medium (Gibco) supplemented with 10% fetal bovine serum (Gibco). #1 coverslips were treated with a 5 µg/mL solution of fibronectin (Sigma) in phosphate-buffered saline (PBS) and let to dry at room temperature. Cells at passage 18 were detached from culture flasks using trypsin, seeded directly onto the fibronectin-treated coverslips, and incubated 24 hours in a 37C/5% $CO_2$ atmosphere until cells adhered and spread on the coverslips. Each sample was fixed for 10 minutes at room temperature in 4% paraformaldehyde in PBS, rinsed twice with PBS, and sealed with a second coverslip.

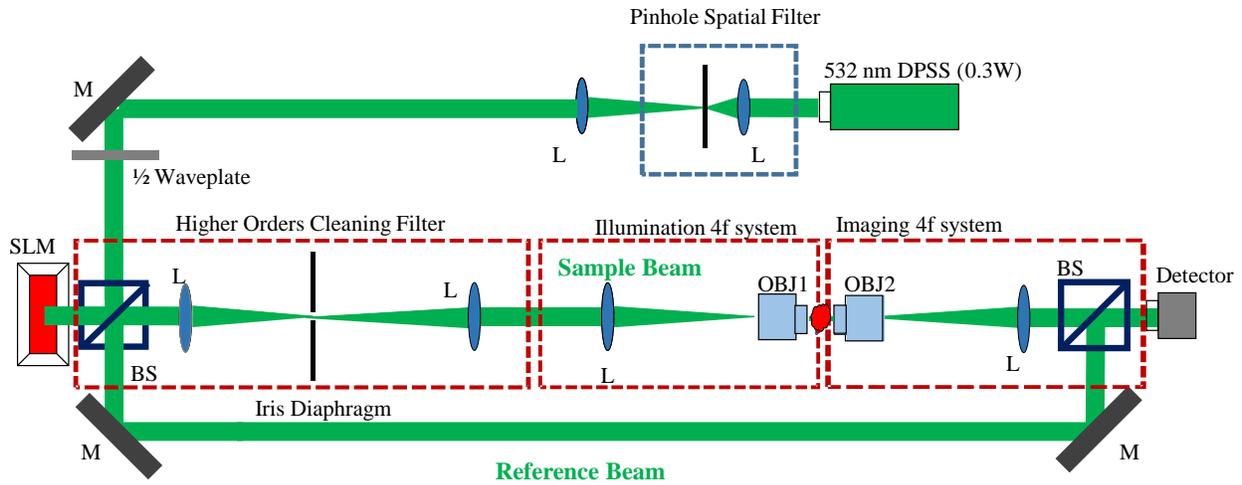

Fig. 1. Experimental tomographic setup. (M: Mirror, L: Lens, OBJ: Objective lens, BS: Beam splitter). Pinhole spatial filter cleans out the beam spatially. The Higher orders cleaning filter removes the unneeded higher orders that might interfere at the image plane on the sample causing image deterioration.

### 3. Computational Algorithms

For 3D refractive index reconstruction, three computational techniques were considered; Rytov, Born, and Radon. ODT was first described by Wolf [1, 2] and refined by Devaney [6]. Like the first order Born approximation, the first order Rytov approximation is also a linearization of the inverse scattering problem but it has been found to yield superior results for biological cells and has been the most commonly used technique for linear ODT [27, 28]. One of the main differences between the Rytov and the Born models is the phase unwrapping that is explicit in the Rytov model [29]. This unwrapped phase is used instead of the field in the inversion formula introduced by Wolf (which we refer to as the Wolf transform). The third technique, the Radon direct inversion based reconstruction [9], is a filtered back-projection reconstruction algorithm that is based on diffraction-free model thus it generates errors when it comes to diffracting objects with spatial variations comparable to the wavelength of light.

In the studied samples (i.e. HCT-116 cells and Panc-1 cells), the accumulated phase from the samples, whose thickness is around 8 μm, exceeds $2\pi$ at some regions depending on the proteins distributions as shown in Fig. 2a, b and thus both Radon and Born fail to reconstruct the 3D refractive index distribution due to considerable diffraction, and high phase accumulation by the sample, respectively. Phase Unwrapping algorithm was used to unwrap the phase [30].

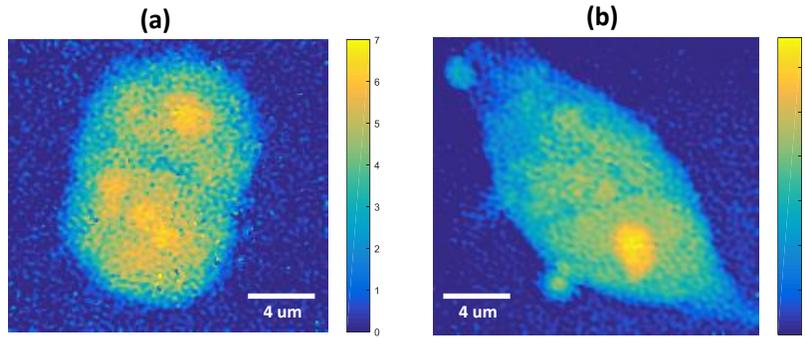

Fig. 2 Unwrapped phase of (a) HCT-116 cell and (b) Panc-1 cell for normal incidence. Phase unwrapping was done using PUMA algorithms [30]. Color bars are in Radians.

As can be seen from Figs. 3 and 4, the Born approximation appears to erroneously estimate the refractive index distribution. Cancer cells usually have a refractive index (RI) of cytoplasm that range between 1.36-1.39 due to excess RNA and protein [7]. As observed in Fig. 3, this index range is probably under-estimated (i.e. around 1.32) due to high phase delay that Born cannot deal with. On the contrary, the Rytov approximations shows better agreement with what is expected from the biology of cells. The estimated index of the cytoplasm is around 1.365 wich is within the expected range. Another interesting feature are the lipids which are composed of fats, sugars and proteins and are characterized by their high proteins concentrations and thus high RI value. This is in agreement with the Rytov reconstructins where we can see bright spots which do not show up in the Born approximation [7].

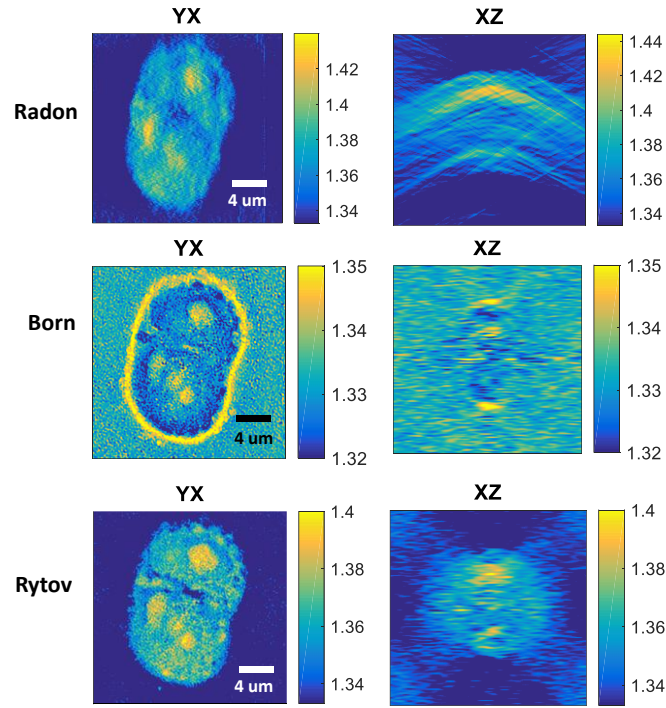

Fig. 3 Different RI Reconstruction of the HCT-116 cell based on Radon, Born, and Rytov techniques.

In Fig. 4, it is obvious how the Born under-estimates the RI value of the nucleus as well where it should have much higher RI than the surrounding media (i.e. water) [7]. This could be because phase unwrapping is not considered and that is why we can see enhanced edges at the boundaries of the cell at the point where the phase wraps while the higher phase is under estimated. However, as can be seen, the RI contrast between nucleus and medium is quite low in case of Born. On the other hand, Rytov agrees with literature where the high RI contrast is clear [7, 31-33].

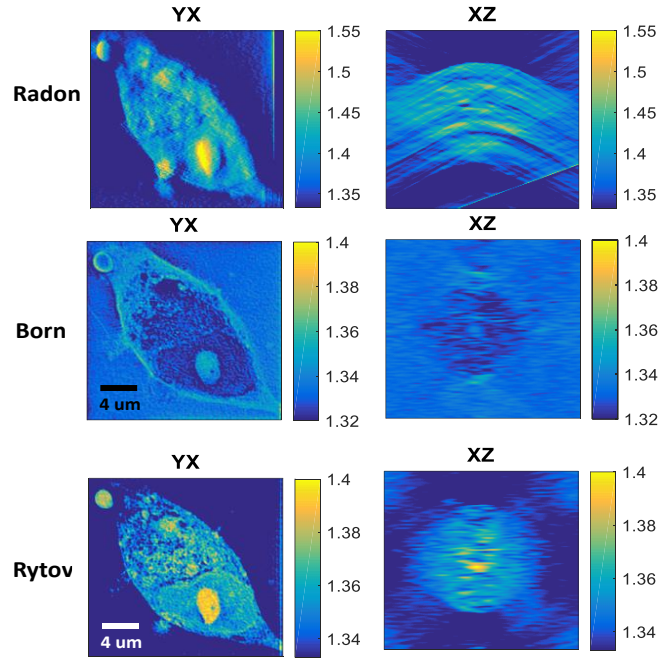

Fig. 4 Different RI Reconstruction of the Panc-1 cell based on Radon, Born, and Rytov techniques.

## 4. Assessment Results

We quantitatively assess the performance of each of the three ODT reconstruction methods in the same experimental set up shown in Figure 1 using digital phase conjugation method we described in Section 2. This is done in two steps; firstly, structured illumination (an image of *Einstein* or *1951 USAF resolution test chart*) was used to phase modulate the incident beam that propagates through the sample and be collected through an interferogram on the sCMOS camera; the wrapped phase of such an image is shown in Fig. 5. In addition, we repeat the exact same measurement by propagating through the media without the cell (clear PBS liquid between two coverslips) by shifting the sample until no cells are in the field of view. This is feasible by controlling the concentration of the HCT-116 and/or Panc-1 inside the PBS to obtain cell-empty regions. The second step in the assessment is done computationally by back-propagating the modulated output (i.e. *Einstein/1951 USAF resolution test chart* modulated with the HCT-116/Panc-1 phase delay) through the reconstructed 3D refractive index (RI) map by using the Lippmann-Schwinger Equation (LSE) model which is known to be accurate even for thick samples and high RI contrasts [24]. By solving an inversion problem (where the inversion of an operator is required) that takes into account transmission and reflection from the scattering media, we were able to efficiently calculate the back-propagated field through the complex media. The LSE method requires a large memory and long processing time (as compared to the beam propagation method for example where reflections are neglected), however this method is more accurate. As we discussed earlier, if the ODT estimate of the 3D index distribution is accurate then we expect to obtain a clean reconstruction of the original structured illumination pattern. Comparison with the original pattern measured from the experiment without the cell gives us a quantitative meaure of the accuracy of the ODT method. Fig. 6 shows the retrieved Einstein and 1951 USAF resolution test chart for the case of Radon, Born, and Rytov approximations as compared to the original field using this procedure.

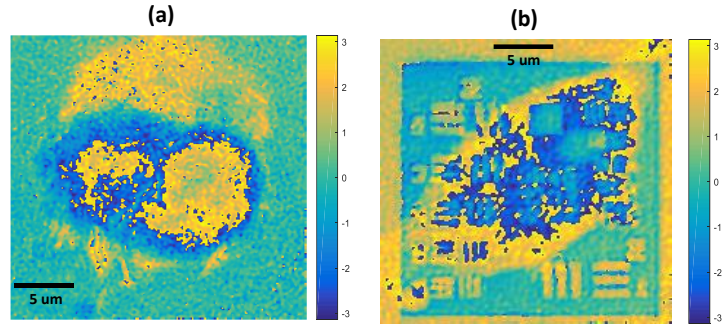

Fig. 5 Wrapped phase of *Einstein/USAF chart* after propagating through the HCT-116/Panc-1 cell.

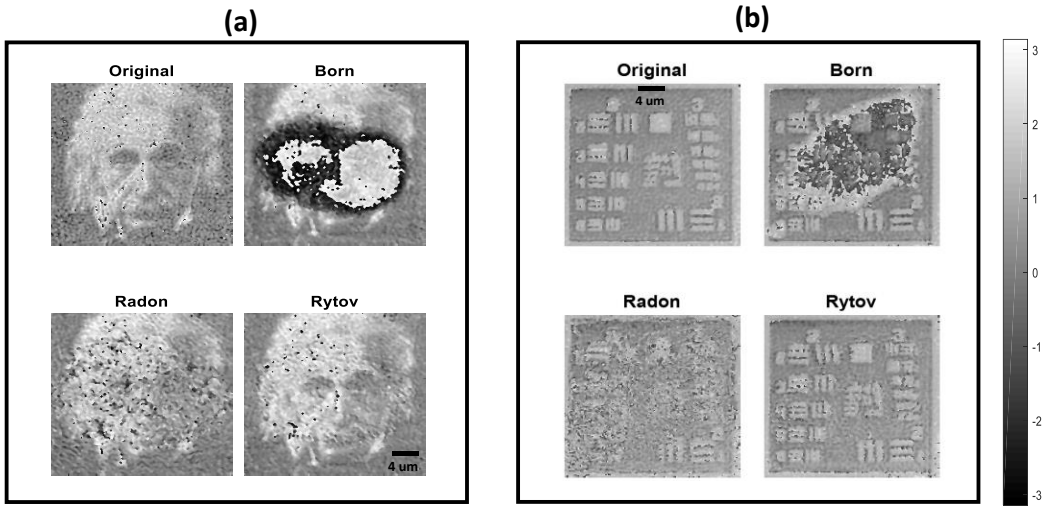

Fig. 6. Retrieved projected fields using Radon, Born, and Rytov for (a) *Einstein* through HCT-116 cell, and (b) *USAF* chart through Panc-1 cell.

To quantify the error, the mean square error (MSE) is measured as follows:

$$Error = MSE(Einstein^{\text{Retrieved}}, Einstein^{Original}) \qquad (1)$$

Assessment was done for reconstructions provided by Radon, Born, and Rytov approximations as shown in Tables 1, and 2. The MSE in the case of Born reconstruction is high as compared to Radon and Rytov (at least 3 times bigger than MSE for Rytov). This is due to the fact that both Radon and Rytov make use of the unwrapped phase whereas the Born reconstruction algorithm is implemented on the complex field. The Radon reconstruction scheme depends on the unwrapped phase, however it ignores diffraction which limits its performance as compared to Rytov which have the best performance by taking into account phase unwrapping, and diffraction.

*Table 1. MSE percentage for Radon, Born and Rytov based Reconstruction techniques for Einstein*

| Radon | Born | Rytov |
|---|---|---|
| 8.83% | 34.73% | 6.39% |

Although the MSE values are changing from one case to another (depending on phase profile, dimensions and diffraction strength), the three reconstructions seems to follow the same trend where Born still have the worst performance as compared to Radon and Rytov as long as phase wrapping occurs.

*Table 2. MSE percentage for Radon, Born and Rytov based Reconstruction techniques for USAF chart*

| Radon | Born | Rytov |
|---|---|---|
| 16.19% | 24.58% | 7.97% |

## 5. Conclusion

We showed how structured illumination can be used to assess the performance of different reconstruction schemes through the use of an SLM for both angular scanning and structured illumination. Having the same experimental setup for angular and structured illumination without the burden of alignment and/or mechanical instabilities, is it possible to evaluate the performance of the different reconstruction algorithms by quantifying the error between different reconstructions based on the retrieved field from the digitally back-propagated output field recorded on the detector using the LSE. This assessment method is useful when imaging biological samples where the ground-truth cannot be known while the reconstructions need to be validated without external reinforcements.